\documentclass[fleqn,twoside]{article}
\usepackage{espcrc3}

\usepackage{graphicx}

\newcommand{\AmS}{{\protect\the\textfont2
  A\kern-.1667em\lower.5ex\hbox{M}\kern-.125emS}}

\title{Cosmic Solutions in the Einstein-Weinberg-Salam Theory and
the Generation of Large Electric and Magnetic Fields\thanks{\it To appear in
the Proceedings of the ICHEP 2002, Amsterdam, 25 July - 31 July, 2002.}}

\author{Yutaka Hosotani\address[Osaka]{Department of Physics, Osaka
University,  Toyonaka, Osaka 560-0043, Japan}, 
Hiroki Emoto\address{High Energy Accelerator Research Organization (KEK),
Tsukuba, Ibaraki 305-0801, Japan}, and
Takahiro Kubota\addressmark[Osaka]}

\newcommand{\beeq}{\begin{equation}}
\newcommand{\eneq}{\end{equation}}
\newcommand{\beqn}{\begin{eqnarray}}
\newcommand{\eeqn}{\end{eqnarray}}

\def\mybig{\displaystyle \strut }

\begin{document}

\begin{abstract}
In the $SU(2)_{L}\times U(1)_{Y}$ standard electroweak theory 
coupled with the Einstein gravity, 
new topological configurations naturally emerge, if the spatial 
section of the universe is globally a three-sphere
($S^3$) with  a small radius.  The $SU(2)_L$ gauge fields wrap the space
nontrivially, producing homogeneous but anisotropic space.  As the universe
expands, large electric and magnetic fields are  produced.  The
electromagnetic field configuration is characterized by the Hopf map.

\hfill OU-HET 417/2002, KEK-TH-842
\vspace{1pc}
\end{abstract}

\maketitle

\section{STANDARD MODEL IN GRAVITY}

It has been firmly established that the standard model of electroweak 
interactions based on the $SU(2)_{L}\times U(1)_{Y}$ gauge symmetry
describes the Nature at low energies, and the evolution of the universe 
is well described by Einstein's theory of gravity.  Given these two facts,
it is natural to ask if there was strong interplay between the Einstein
gravity  and electroweak interactions in the early universe.  We report that
indeed such strong interplay may have existed, provided  the universe is
spatially closed.\cite{EWS}.

\section{TOPOLOGICAL CONFIGURATIONS}

The Lagrangian density  is given by
\[
{\cal L} = \frac{1}{16\pi G} (R-2\Lambda )
-\frac{1}{4} F^{a}_{\mu \nu}F^{a\mu \nu} 
-\frac{1}{4} G_{\mu \nu}G^{\mu \nu}
\]
\[  
\hskip 2.cm 
- (D_{\mu }\Phi)^{\dag} (D^{\mu}\Phi ) 
- \lambda \Big(\Phi ^{\dag}\Phi - \frac{v_0^2}{2} \Big)^{2}  
\]where $D_{\mu }\Phi = (\partial_{\mu}- i \frac{g}{2} \,
\tau^{a}{A_{\mu}^{a}} -i \frac{g'}{2} B_{\mu})\Phi$.  $\Lambda$ is the
effective cosmological constant at the time the universe was wrapped with
nontrivial field configurations discussed below.

Recall that the metric of an unit three-sphere $S^3$ can be written as
$d{\Omega_3}^2 = \sum_{j=1}^3 \sigma^j \otimes \sigma^j$ where
the $SU(2)$ Maurer-Cartan 1-forms $\sigma^j= {\sigma^j}_\mu dx^\mu$
satisfy $d\sigma^j = \epsilon^{jkl} \sigma^k \wedge \sigma^l$.
The metric of the spacetime we consider is
\[
ds^2 = - e^0 \otimes e^0 + \sum_{j=1}^3 e^j \otimes e^j
\]
\beeq
e^0 = dt ~,~  e^j = a_j(t) \sigma^j ~.
\eneq
It gives a homogeneous but anisotropic space.  The $SU(2)$ and
$U(1)$ gauge fields and the Higgs fields take
\[
A = \frac{1}{2g} \, 
\sum_{j=1}^3  f_j(t) \sigma^j \tau^j ~~,  ~~
\Phi = \mybig {1\over \sqrt{2}}     \pmatrix{ 0 \cr v(t) \cr}
\] 
\beeq
B = h(t) \sigma^3 
\eneq
where $\tau^j$'s are Pauli matrices.  The Ricci tensors and
the energy-momentum tensors in these tetrads become diagonal and independent
of  the spatial coordinates.  Both the Einstein equations and field
equations are consistently solved.  

The spatially isotropic universe where all scale factors $a_j(t)$'s
are equal to each other is not permitted in the presence of the $U(1)$ gauge
interactions.  It still is legitimate to set $a_1=a_2$ and 
$f_1=f_2$ when the Higgs field and the $U(1)$ gauge field are aligned 
as  in (2).  The
ansatz of this type has been investigated before in the pure
Einstein-Yang-Mills theory in
\cite{Hosotani}.

\section{LOCAL MINIMUM OF THE POTENTIAL}

The configuration (2) is topological in the sense that the $SU(2)$
gauge fields wrap the space nontrivially if $a_j(t)$'s are sufficiently
small.  However it is not  absolutely stable as it can be continuously 
deformed to the trivial configuration $A=B=0$.  
To see how it may become important we insert the ansatz (2) into the
potential:
\[
V = {\lambda\over 4} (v^2 - v_0^2)^2
+ {v^2 \over 8}  \left\{ {2 f_1^2\over a_1^2}
  + {(f_3 - g' h)^2 \over a_3^2} \right\}  
\]
\beeq
\hskip .8cm 
+ { (2 f_3 - f_1^2)^2 \over 2 g^2 a_1^4}
         + {2 f_1^2 (f_3 - 2)^2 \over 2 g^2 a_1^2 a_3^3} 
       + {2h^2\over a_1^4} ~.
\eneq
The global minimum of the potential is always located at
$v=v_0$, $f_1=f_3=h=0$.  When the scale factors $a_1$ and $a_3$ are 
sufficiently small, there appears a new local minimum located at
$1.5 < f_1, f_3 \le 2$ and $0 \le v < v_0$.

It could well be that the very early universe was settled
near the local minimum of the potential.  The local minimum has
positive energy density so that the Einstein equations drive  the
universe to expand.  As $a_j$'s become larger, the terms of $O(a^{-2})$ and
$O(a^{-4})$ in $V$ become less important.  The barrier 
separating the local minimum from the global minimum quickly 
disappears.

\section{ELECTROMAGNETIC FIELDS}

After the barrier disappears, the field configuration starts
to roll down the hill of the potential.  It is in this period of
the evolution of the universe that some of interesting physical
consequences  emerge.  One salient feature of the electroweak
theory is that there remains the electromagnetic $U(1)_{\rm EM}$ gauge
invariance after the Higgs field develops a nonvanishing vacuum expectation
value, i.e.\
$v \not= 0$.  The electromagnetic fields  are given by
\[
\hskip 1cm
F_{\rm EM} = {\dot h_{\rm EM}\over a_3}  e^0 \wedge e^3
+  {2 h_{\rm EM}\over a_1^2 }  e^1 \wedge e^2  ~~,
\]
\beeq
\hskip 1cm
h_{\rm EM} (t) =  {1\over \sqrt{g^2 + g'^2}} 
\left\{ g h + g' {f_3\over g} \right\} ~~.
\eneq

\begin{figure}[h]
\begin{center}
\hskip 0.cm 
\includegraphics[width=7.0cm]{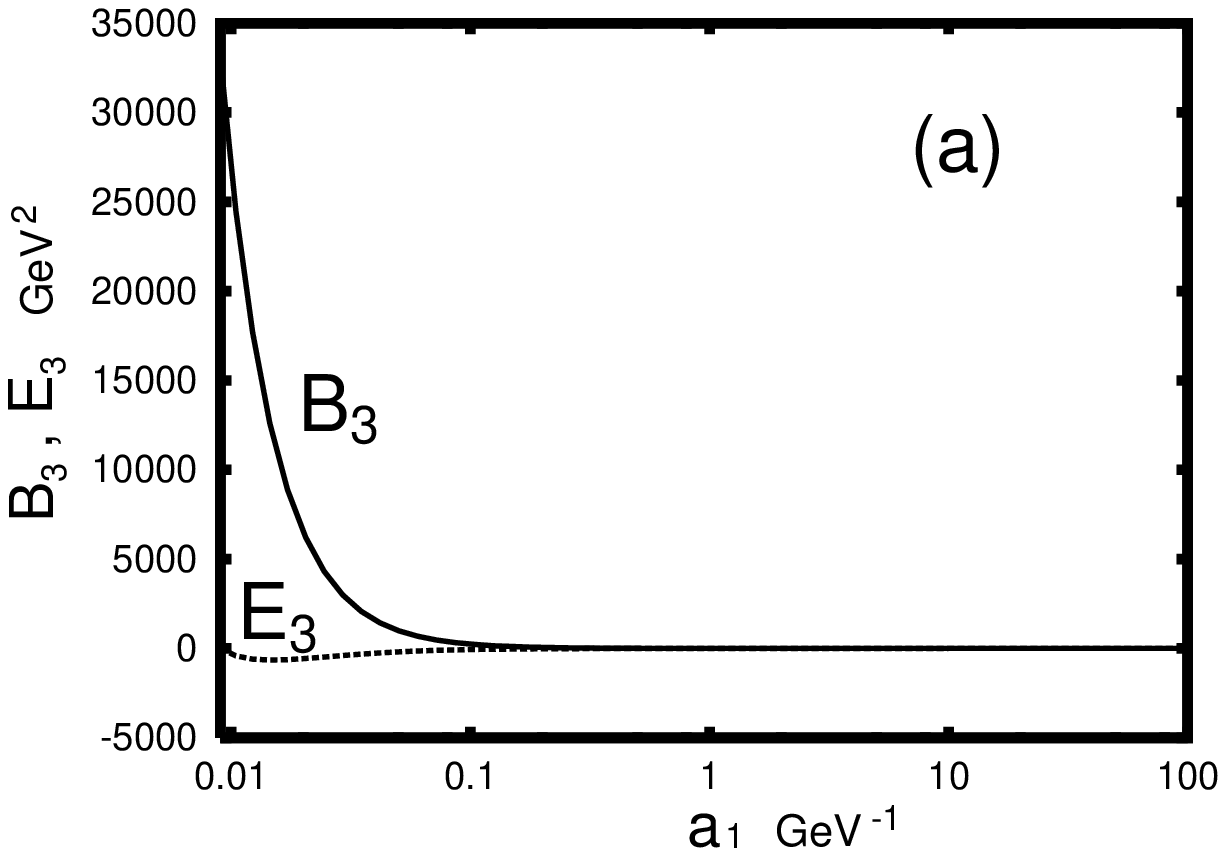}
\includegraphics[width=7.0cm]{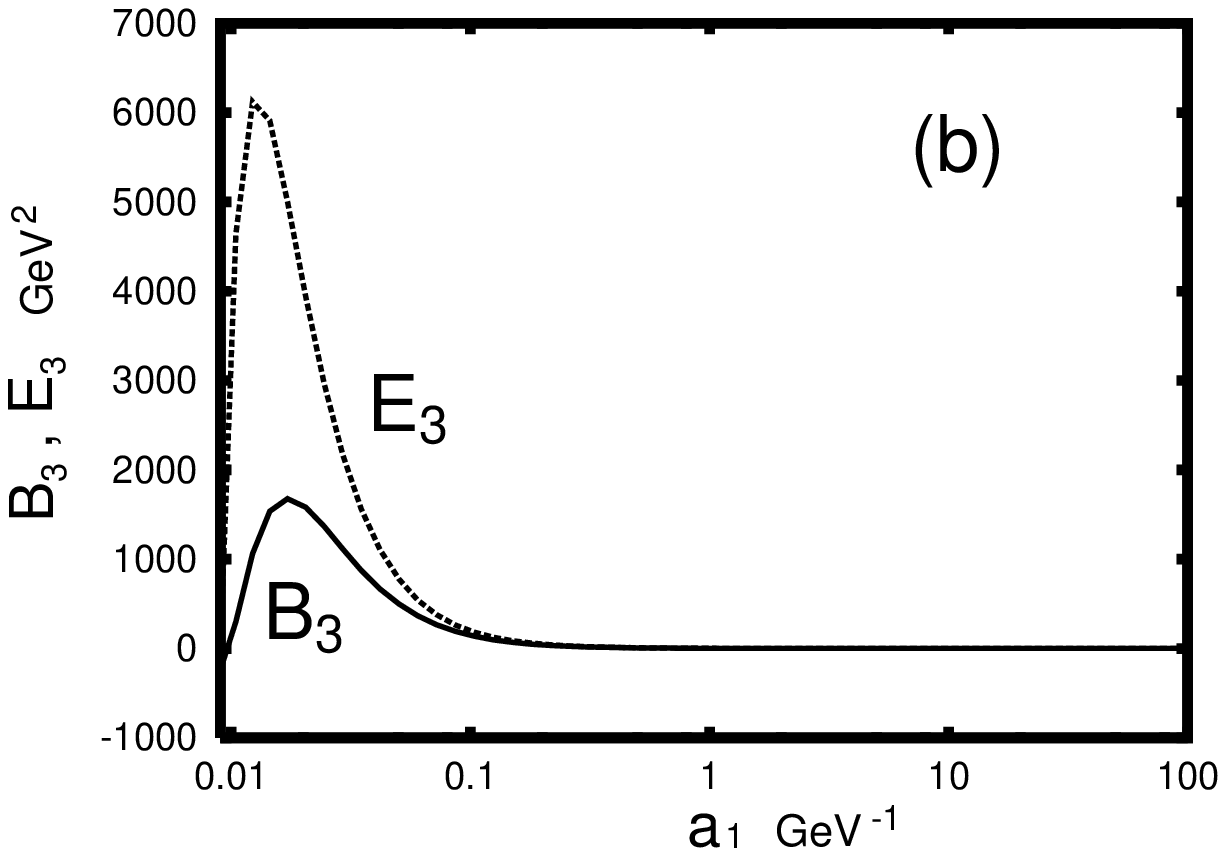}
\vskip -.5cm
\caption{Generated electromagnetic fields as functions of 
$a_1(t)$.  $\Lambda= 1.0 \times 10^5 {\rm GeV}^2$ and
$a_j(t_0) = 9 \times 10^{-3} {\rm GeV}^{-1}$.  (a) The fields
started from the local minimum of the potential at $t_0$. 
(b) Initially $B_3=E_3=0$.}
\end{center}
\end{figure}

In the potential (3), the terms of $O(a^{-4})$  quickly become
irrelevant as $a_j$'s increase.  The terms of $O(a^{-2})$  are
independent of
$h_{\rm EM}$, reflecting the $U(1)_{\rm EM}$ invariance.
In other words, the potential has approximately flat direction along
$f_1=f_3-g' h=0$.  During the expansion the field configuration 
heads for this shallow valley of the potential,
but not necessarily for the absolute minimum.  $h_{\rm EM}$ approaches
a nonvanishing value, producing large electromagnetic
fields.  Typical behavior is depicted in fig.\ 1.

A typical value for the generated magnetic fields  is about
$(50 {\rm GeV})^2$.  It dies away as $a(t)^{-2}$ as the universe further
expands.
The final value of $h_{\rm EM}$ depends on
the initial condition, but the fact $h_{\rm EM}(t=\infty) \not= 0$
does not.  In fig.\ 2 the final values of $h_{\rm EM}$ and $f_3$ are plotted
as functions of the initial value of $h_{\rm EM}$.  The figure shows 
how generic it is to generate large electromagnetic fields.

\section{FIELDS AS THE HOPF MAP}

As is evident in eq.\ (4), the magnitude of the generated electromagnetic
fields is independent of the spatial position.   Both the electric
and magnetic fields are  in the $e^3 = a_3 \sigma^3$ direction where
$\sigma^3$ varies as the spatial position.
They define smooth, regular vector fields without any singularity.

The space is topologically isomorphic to $S^3$.  At each instant $t$,
the vector $\vec E (x)$ or $\vec B (x)$  is on $S^2$ specified in the field
space by $|\vec E|$ or $|\vec B |=$constant.  
The vector field defines a map from $S^3$ onto $S^2$.
It is nothing but the Hopf map.  

This is how it becomes possible to
have spatially homogeneous universe with nonvanishing $U(1)$ gauge
field strengths.  If we were living in two-dimensional space,
the gauge field strengths with constant magnitude would necessarily
induce singularities, either in the form of sources/sinks or
vortices.

\section{A COLD ERA OF THE EARLY UNIVERSE}

When does the field configuration (2) become important in the 
history of the universe?   In the standard scenario of the early universe,
the temperature  effect plays an important role.   It modifies 
the effective potential and the time-evolution.   What we have in
mind here
 is an era preceding the hot universe which continuously  evolves
to the current universe.  Suppose that at one instant the universe was very
small and cold, and the gauge and Higgs fields assumed the nontrivial
configuration under discussions. Driven by an effective cosmological
constant the universe underwent inflation, and 
large electromagnetic fields were generated. Eventually the
inflation stopped and the universe was reheated to the temperature
about $(\Lambda/8\pi G)^{1/4}$.  The universe   continued to expand by
the radiation dominance since then.

\begin{figure}[tbh]
\begin{center}
\hskip 0.cm 
\includegraphics[width=7.cm]{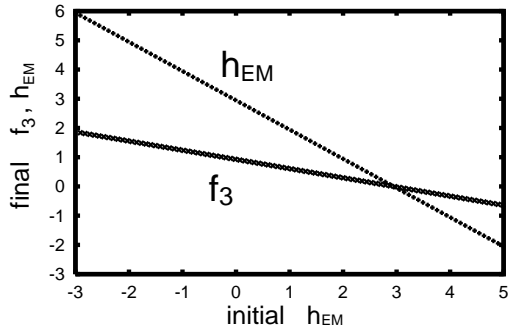}
\vskip -.5cm
\caption{The final $h_{\rm EM}$ and $f_3$ v.s.\ the initial $h_{\rm EM}$. 
$\Lambda = 1 \times 10^6$ GeV$^2$ and $a_j(t_0)= \sqrt{3/\Lambda}$.}
\vglue -.5cm
\end{center}
\end{figure}

When  large electromagnetic fields were generated, quarks and leptons 
must have been copiously created.\cite{Gibbons}  As the field
configuration was not invariant under $CP$, there may have been asymmetry
in the baryon and lepton numbers as well.  Such asymmetry could be washed
out in the later stage of the  evolution of the universe, but a part of it
may have survived.

In this report we have assumed the homogeneity of the space.
In reality inhomogeneity has to be generated either classically or
quantum mechanically.  The generated magnetic flux, which is taken
to be homogeneous in our ansatz (2), may be squeezed in space.  Could such
magnetic flux have remnant in the present universe?\cite{Barrow}  We do not
have an answer at the moment.

\end{document}